\documentclass[aps,prl,reprint,superscriptaddress]{revtex4-2}

\usepackage{graphicx}
\usepackage{dcolumn}
\usepackage{bm}
\usepackage{color}
\usepackage{xcolor}
\usepackage{epstopdf}
\usepackage{gensymb}
\usepackage{ulem}
\usepackage{float}
\usepackage{xr}
\usepackage{sidecap}
\usepackage{mathrsfs}
\usepackage{amsmath}
\usepackage{amsthm}
\usepackage{enumerate}
\usepackage{soul}
\usepackage{hyperref}
\usepackage{amssymb}
\usepackage{units}
\usepackage{afterpage}
\usepackage{lineno,hyperref}
\usepackage{comment}
\usepackage{graphicx}

\begin{document}
\preprint{APS/123-QED}

\title{Beyond von Mises Truss Models: Emergent Bistability in Mechanical Metamaterials\\}

\author{Md Nahid Hasan}
\affiliation{Department of Mechanical Engineering, University of Utah, Salt Lake City, UT 84112, USA}
\affiliation{Department of Mechanical Engineering, Montana Technological University, Butte, MT 59701, USA}

\author{Taylor E. Greenwood}%
\affiliation{Department of Mechanical Engineering, University of Utah, Salt Lake City, UT 84112, USA}
\affiliation{Department of Mechanical Engineering, Pennsylvania State University, University Park, PA 16802, USA}

\author{Sharat Paul}%
\affiliation{Department of Mechanical Engineering, University of Utah, Salt Lake City, UT 84112, USA}


\author{Bolei Deng}%
\affiliation{Guggenheim School of Aerospace Engineering, Georgia Institute of Technology, Atlanta, GA 30332, USA}

\author{Qihan Liu}%
\affiliation{Department of Mechanical Engineering and Materials Science, University of Pittsburgh, Pittsburgh, PA 15213, USA}

\author{Yong Lin Kong}
\email{kong@rice.edu}
\affiliation{Department of Mechanical Engineering, University of Utah, Salt Lake City, UT 84112, USA}
\affiliation{Department of Mechanical Engineering, Rice University, Houston, TX 77005, USA}
\affiliation{Rice Advanced Materials Institute, Rice University, Houston, TX 77005, USA}
\author{Pai Wang}
\email{pai.wang@utah.edu}
\affiliation{Department of Mechanical Engineering, University of Utah, Salt Lake City, UT 84112, USA}


\begin{abstract}
We observe and analyze the phenomenon of bistability emergent from cooperative stiffening in hyper-elastic metamaterials. Using experimental and numerical results of identical geometric designs, we show evidence that a single unit is unistable while combining two units can result in bistability. Our study demonstrates that the von Mises truss model cannot describe such emergent behavior. Hence, we construct a novel and simple analytical model to explain this phenomenon.
\end{abstract}
\maketitle
\textit{Introduction:} 
\begin{figure}[t!]
\centering
\includegraphics[width=\linewidth]{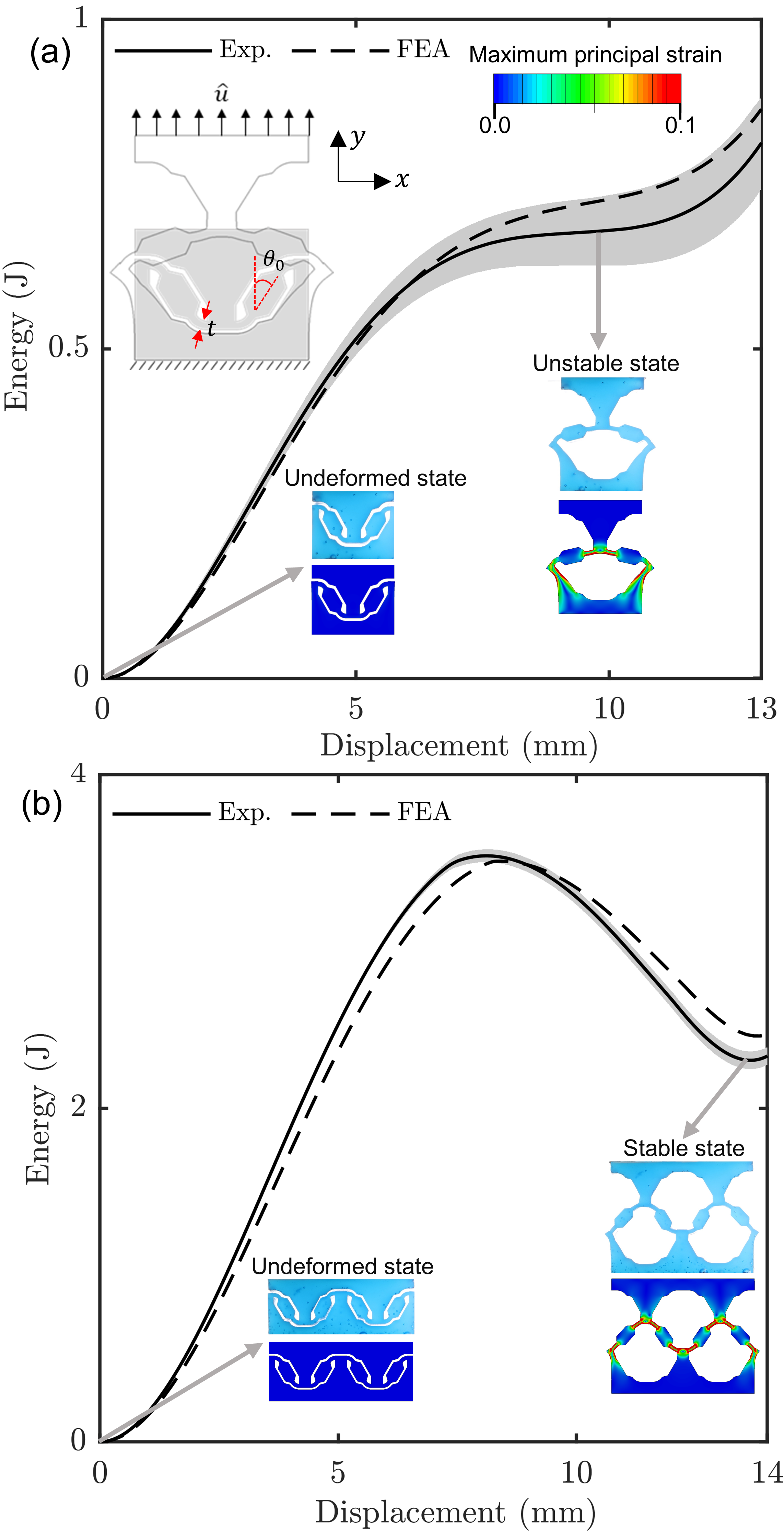}
\caption{Energy landscape of mechanical metamaterials: (a) The unistable unit cell, depicted in both contracted and expanded states with gray geometry. This is complemented by a comparison between experimental results and simulations, highlighting the absence of an energy barrier. (b) The bistable two-unit cell features both experimental and simulated energy landscapes, showing the presence of an energy barrier. The shaded areas denote the standard deviation within five measurements of energy-displacement responses for the single and two-unit cells, whereas the solid lines represent the average values.}
\label{fig:Fig1}
\end{figure}
Multistable mechanical metamaterials are crucial for applications requiring adaptability~\cite{greenwood2025soft, giri2021controlled,yang2023shape,xiu2022topological,rahman2025reprogrammable} and reconfigurablity~\cite{tao20204d, wu2023topological,zhang2023bistable} due to their capability of morphing between different shapes and states. Such applications include soft robotics~\cite{chi2022bistable, chen2018harnessing,carton2025bridging}, energy absorption~\cite{shan2015multistable, chen2023elastic,yue2024flexibly}, and actuation devices~\cite{chi2022bistable}. Theoretical and computational analyses of these metamaterials often rely on examining a single unit cell with periodic boundary conditions (PBC), assuming an infinite array of repeating unit cells, to understand their mechanical behavior~\cite{che2017three,zhang2024characteristic,restrepo2015phase,jeon2022synergistic,wu2023tunable}. However, PBC may fail to account for the mechanical responses arising from interactions between unit cells, particularly by not considering the finite size effects~\cite{liu2024suppression,maraghechi2020experimental,wang2024topological}. 
A transition from unistable to bistable behavior may occur in systems with two or more interconnected units~\cite{peng2024programming}. Current analytical approaches, such as the von Mises truss model~\cite{chen2018harnessing,yang2023shape,barbarino2013bi,wan2024finding,chen2017integrated,liu2023novel,de2023static,falope2021snap,lietard2024exploiting,barbarino2013bi}, fall short of capturing interunit interactions. Recognizing the limitations of PBCs and the widely used von Mises truss models, we aim to develop a new model to capture, both qualitatively and quantitatively, the behavior of mechanical metamaterials in finite-size samples.

In this letter, building on our recently published design and observations~\cite{greenwood2025soft}, we investigate the interunit interaction of multistable mechanical metamaterials, moving beyond the intrinsic assumptions of PBCs. The experimental and computational observations reveal a transformation from the unistable behavior of a single-unit sample to the bistability of a two-unit sample, highlighting the significance of interactions between neighboring units and their collective stiffness. We introduce new theoretical insights on bistability in metamaterials, challenging the currently prevailing von Mises truss model. The research emphasizes the value of studying finite systems, enabling a clearer analysis of stability transitions that are often overlooked in unit-cell analyses. The findings highlight the need to reevaluate existing models and theories in light of the emergent behaviors of mechanical metamaterials, facilitating the translational impact of engineering applications. 

\textit{Observation of emergent bistability:}
We initiate our study by analyzing the numerical predictions and experimental observations. Figure~\ref{fig:Fig1}(a) shows the quasi-static response of a single-unit cell.  The inset details the geometric parameters of the unit cell, including the thickness of the inclined beam (\(t = 0.96 \, \text{mm}\)). 
The angle between the rigid section and the vertical direction is \(\theta_0 = 30^\circ\) (see supplemental
material~\cite{SI} for more details). The metamaterial is cast from high performance platinum-cured liquid silicone (Dragon Skin 30) that is nearly incompressible (Poisson ratio \(\nu \approx 0.495\)), with an initial Young's modulus of \(Y = 0.74 \pm 0.07\) MPa~\cite{DragonSkin30,ranzani2015bioinspired} and a density of (\(\rho = 1080 \, \text{kg/m}^3\)). We conduct experiments on two samples at room temperature: one is a single-unit sample, and the other is a sample consisting of two unit cells.
While the bottom of each sample is fixed,  a uniaxial tensile displacement loading at 10 mm/min is applied to the top surface.
The energy landscape of the single-unit sample - solid black curve in Fig.~\ref{fig:Fig1}(a) - confirms the unistable behavior without any energy barrier. In contrast, we observe bistability in the two-unit sample, shown as the solid-black curve in Fig.~\ref{fig:Fig1}(b), where we observe a clear energy barrier. We perform five measurements for each sample, and the gray-shaded regions in Fig.~\ref{fig:Fig1} indicate the standard deviation across these trials. 

To computationally capture this observation, we conduct finite element analyses (FEA) on the commercial platform \textsc{abaqus/standard} with the dynamic implicit step with the quasi-static option activated~\cite{hua2020parameters,zhang2021novel,wu2022design,li2022design}. We model the metamaterials using two-dimensional (2D) plane strain elements (CPE4H) and employ the Neo-Hookean model to characterize the hyperelastic material properties.
Similar to our experimental procedures, we evaluate the static response by applying a controlled vertical displacement to induce the transition from a contracted to an expanded state, with the base fully constrained. The dashed curves in Fig.~\ref{fig:Fig1} depict the simulation results, which show good agreement with our experimental observations. Both single-unit and two-unit samples undergo highly nonlinear, large deformations during the tensile loading process. As shown in the insets of Fig.~\ref{fig:Fig1}, we also present the sample shapes at key states. The experimental and numerical results in terms of the deformed geometries also show good agreement with each other, while the numerical results provide additional information on localized strain concentrations.

\begin{figure}[t]
    \centering
\includegraphics[width=1.0\linewidth]{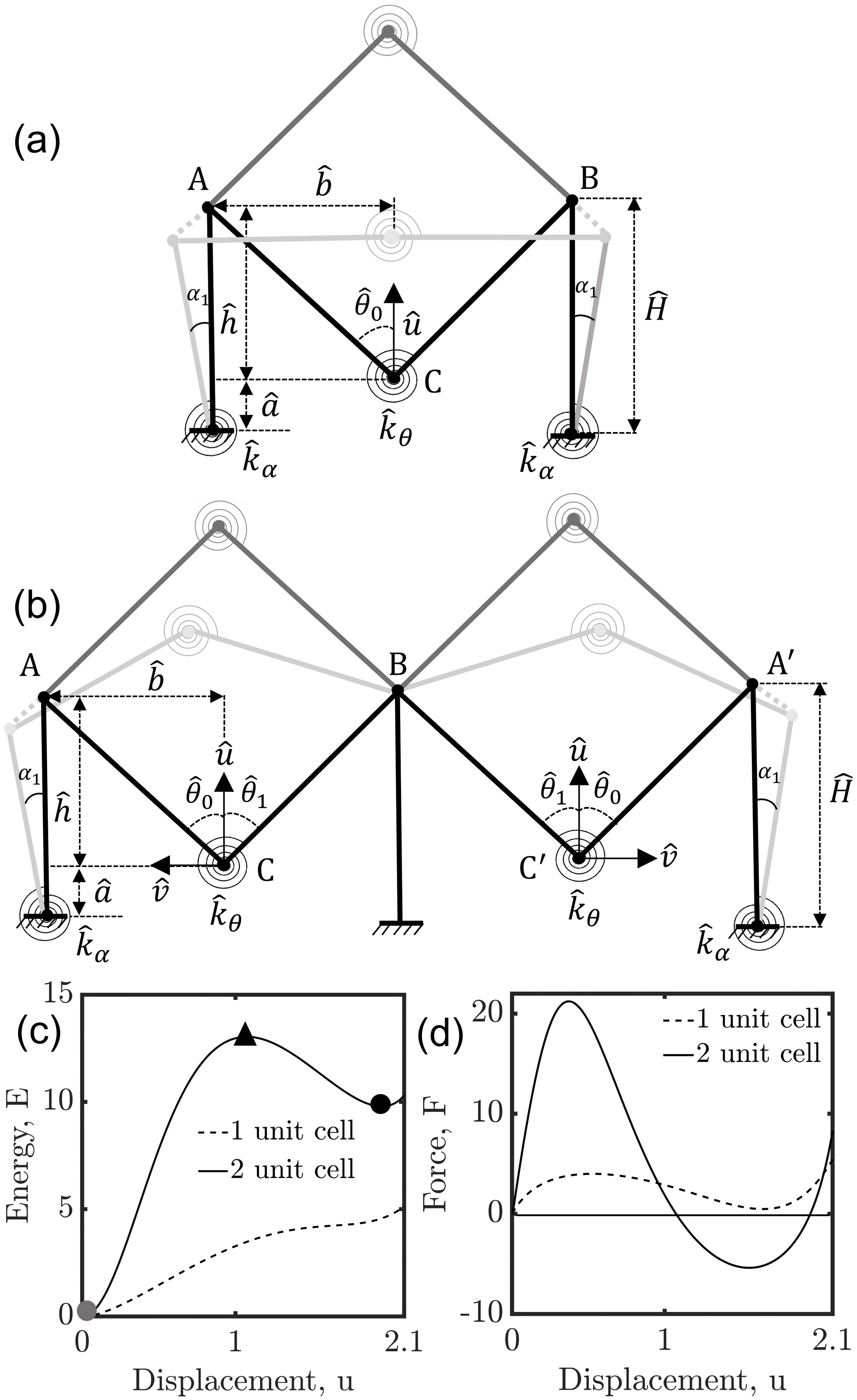}
    \caption{Analytical realization of the emergent bistability: (a) Analytical discrete model of a single-unit cell. (b) Discrete model for a two-unit cell assembly. (c) The energy landscape clearly shows forward and reverse transition energy barriers, \(\Delta E_{\mathrm{f}} = E_{\mathrm{max}} - E_{\mathrm{w_1}}\) and \(\Delta E_{\mathrm{r}} = E_{\mathrm{max}} - E_{\mathrm{w_2}}\), respectively, for the two-unit cell. In contrast, there is no energy barrier for the single-unit cell. Gray circles \textcolor{gray}{$(\bullet$)}, black triangles $(\blacktriangle)$, and black circles $(\bullet)$ denote the energy at the first stable state, the energy at the peak of the barrier, and the second stable state \(E_{\mathrm{w_1}}\), \(E_{\textrm{max}}\), \(E_{\mathrm{w_2}}\), respectively. (c) Force-displacement curves illustrating the transition from unistability in a single-unit cell to bistability in a two-unit cell system, with parameters \(k_{\alpha}=22\), \ \(h=1.07\), \(a=0.39\) and \(H=h+a\).}
    \label{fig:Fig2}
\end{figure}

\textit{Analytical discrete model:} To understand the fundamental mechanisms behind the emergent bistability from cooperative interactions between two unistable unit cells, we develop two analytical discrete models in dimensionless form for single-unit and two-unit cells. Figure~\ref{fig:Fig2}(a) shows a single unit cell consisting of two inclined rigid bars, \( \text{AC} \) and \( \text{BC} \). Each bar, has an initial length of \( \hat{L}_0 = \sqrt{\hat{b}^2 + \hat{h}^2} \), inclines at an angle \( \hat{\theta}_0=\text{atan2}\left(\hat b,\hat h\right) \) from the vertical, where the function \textrm{`atan2'} is the four-quadrant inverse tangent~\cite{MathWorks_atan2}. Endpoints A and B of the two inclined bars are separated by a width of \( 2\hat{b} \), with point C positioned at an apex height of \( \hat{h} \). A rotational spring at point C, characterized by a stiffness of \(\hat{k}_{\theta}\), provides bending resistance. Additionally, vertical sidebars at points A and B, each with a height of \(\hat{H} = \hat{h} + \hat{a}\) and a rotational stiffness of \(\hat{k}_{\alpha}\), support the bars AC and BC. The sidebars are also fixed at the base.
We apply a vertical control displacement \(\hat u\) at point C, inducing angular inclinations \( \alpha_1 \) in the sidebars. We normalize the model's dimensions relative to \( \hat{b} \) and the stiffness and energies by \( \hat{k}_{\theta} \) by setting the values of \( \hat{b} \) and \( \hat{k}_{\theta} \) to 1, resulting in dimensionless parameters. This normalization yields \( u = \frac{\hat{u}}{\hat{b}} \), \( h = \frac{\hat{h}}{\hat{b}} \), \( H = \frac{\hat{H}}{\hat{b}} \), \( a = \frac{\hat{a}}{\hat{b}} \), \( E_1 = \frac{\hat{E}_1}{\hat{k}_{\theta}} \), \( E_2 = \frac{\hat{E_2}}{\hat{k}_{\theta}} \), and \( k_{\alpha} = \frac{\hat{k}_{\alpha}}{\hat{k}_{\theta}} \) (see supplemental material~\cite{SI} for detailed derivation). 
The dimensionless total potential energy of the single-unit cell is denoted by:
\begin{equation}\label{Eq:eq1}
E_1 =\frac{1}{2}\left( 2 \Delta \theta_0 \right)^2 + k_{\alpha} (\alpha_1)^2,
\end{equation}
where \(\Delta \theta_0 = (\theta_0' - \theta_0)\) represents the change in the inclination angle of the rigid bar AC (and, by symmetry, also BC) resulting from the displacement \( u \), and \( \alpha_1 \) is the tilting angle of the two rigid vertical sidebars. Here, the deformed angle of the bar AC (and, by symmetry, also BC) is given by \(\theta_0' = \text{atan2}\left(1 + H \sin \alpha_1, h - H - u + H \cos \alpha_1\right)\). Inspecting the kinematics, we can obtain 

\begin{equation} \label{Eq:eq2}
\begin{aligned}
\Delta \theta_0 =& \text{atan2}\left(1 + H \sin \alpha_1, h - H - u + H \cos \alpha_1\right) \\
&- \text{atan2}\left(1, h\right)
\end{aligned}
\end{equation}
Also, we have
\begin{equation}\label{Eq:eq3}
\sqrt{1 + h^2} = \sqrt{(1 + H \sin \alpha_1)^2 + (h - H - u + H \cos \alpha_1)^2},
\end{equation}
which conserves the length of the inclined rigid bars AC and BC. 

Figure~\ref{fig:Fig2}(b) illustrates the discrete model of a two-unit cell system by adding a second identical unit to Fig.~\ref{fig:Fig2}(a). The symmetry about the middle vertical bar introduces the equal and opposite horizontal translation $v$ of the two points C and \(\textrm{C}'\). Figure~\ref{fig:Fig2}(b) also shows the initial inclination angle of bars AC (and, by symmetry, also \(\textrm{A}'\textrm{C}'\)) and \(\textrm{B}\textrm{C}\) (by symmetry, also \(\textrm{B}\textrm{C}'\)) are denoted  \( {\theta}_0={\theta}_1=\text{atan2}\left(1, h\right) \) in the undeformed configuration. 
Due to prescribed vertical displacement \(u\), inclination \(\alpha_1\) of the sidebars, and horizontal translation \(v\), the deformed inclination angles of the bar AC 
and  \(\textrm{B}\textrm{C}\) 
are denoted as \(\theta_0' = \text{atan2}\left(1 + H \sin \alpha_1-v, h - H - u + H \cos \alpha_1\right)\) and \(\theta_1' = \text{atan2}\left(1 + v, h-u\right)\) respectively. Therefore, the dimensionless total potential energy of the two-unit cell is denoted as: 
\begin{equation} \label{Eq:eq4}
E_2 = \left( \Delta \theta_0+\Delta \theta_1 \right)^2+k_{\alpha} (\alpha_1)^2,
\end{equation}
where \(\Delta \theta_0 =(\theta_0' - \theta_0)\) denotes the change of angle of the bars AC (by symmetry, also \(\textrm{A}'\textrm{C}'\)) and \(\Delta \theta_1 = (\theta_1' - \theta_1)= (\theta_1' - \theta_0)\) \ denotes the change of angle \(\textrm{B}\textrm{C}\) (by symmetry, also \(\textrm{B}\textrm{C}'\)) (see supplemental
material~\cite{SI} for detail derivation) :
\begin{align} \label{Eq:eq5}
\Delta \theta_0 &= \Bigg( \text{atan2}\left(1 + H \sin \alpha_1-v, h - H \right. \nonumber \\
&\quad \left. - u + H \cos \alpha_1\right) - \text{atan2}\left(1, h\right) \Bigg),
\end{align}
\begin{equation} \label{Eq:eq6}
\Delta \theta_1 = \text{atan2}(1+v, h - u ) - \text{atan2}(1, h).
\end{equation}
\( \alpha_1 \) is a function of \( u \), maintaining the dimensionless geometric constraints: The length of each rigid bars \( \textrm{A}\textrm{C} \), \( \textrm{A}'\textrm{C}' \), \( \textrm{B}\textrm{C} \), and \( \textrm{B}\textrm{C}' \), does not change:
\begin{equation}\label{Eq:eq7}
\begin{split}
&\sqrt{1 + h^2}\\
= &\sqrt{(1 + H \sin \alpha_1 - v)^2 + (h - H - u + H \cos \alpha_1)^2} ,
\end{split}
\end{equation}
\begin{equation}\label{Eq:eq8}
\sqrt{1 + h^2} = \sqrt{(1+v)^2 + (h-u)^2}, 
\end{equation}
We use MATLAB's \texttt{fsolve} function along with the arc-length method~\cite{VasiosArcLength} to numerically solve Eqs.~\eqref{Eq:eq1} and~\eqref{Eq:eq4}, which account for length conservation as defined for the single-unit and two-unit cells in Eqs.~\eqref{Eq:eq3}, \eqref{Eq:eq7}, and~\eqref{Eq:eq8}. Under the prescribed displacement \(u\), we numerically evaluate Eq.~\eqref{Eq:eq3} to determine the values of \(\alpha_1\) for the single-unit cell and both \(\alpha_1\) and \(v\) for the two-unit cell, based on the constraints of Eqs.~\eqref{Eq:eq7} and~\eqref{Eq:eq8}. We then use the values of \(\alpha_1\) to obtain the energy landscape for a single-unit cell by solving Eq.~\eqref{Eq:eq1}. Similarly, for a two-unit cell, we use the values of both \(\alpha_1\) and \(v\) to get the energy landscape from Eq.~\eqref{Eq:eq4}. The parameters are set as follows: \(k_{\alpha} = 22\), \(h = 1.07\), \(a = 0.39\), and \(H = h + a\) for solving Eqs.~\eqref{Eq:eq1} and~\eqref{Eq:eq4}. Figure~\ref{fig:Fig2}(c) displays the energy landscapes for the single-unit cell with a dashed line and the two-unit cell with a solid line. It clearly demonstrates a distinct energy barrier for two-unit cells and no energy barrier for single-unit cells, highlighting the transitions between stable states in the two-unit cell configuration. We define the forward transition energy barrier for the two-unit cell as:
\begin{equation}\label{Eq:eq9}
    \Delta E_{\mathrm{f}} = E_{\mathrm{max}} - E_{\mathrm{w_1}},
\end{equation}
where $E_{\mathrm{max}}$ is the energy at the peak of the barrier (black triangle in Fig.~\ref{fig:Fig2}(c)), and $E_{\mathrm{w_1}}$ is the local minima of the first stable state (gray circle in Fig.~\ref{fig:Fig2}(c)). Similarly, the reverse transition energy barrier is defined as:
\begin{equation}\label{Eq:eq10}
    \Delta E_{\mathrm{r}} =  E_{\textrm{max}} - E_{\mathrm{w_2}},
\end{equation}
with $E_{\mathrm{w_2}}$ representing local minima of the second stable state (black circle in Fig.~\ref{fig:Fig2}(c)). Thus, Fig.~\ref{fig:Fig2}(c) confirms that our analytical discrete network model successfully captures the observations of emergent bistability.

Figure~\ref{fig:Fig2}(d) displays the force-displacement curves for single-unit and two-unit cells. With the prescribed vertical displacement \(u\), the reaction force \(F\) arises, which we analyze by taking the derivative of the total potential energy for both cell configurations. This relationship is described by the following equation:
\begin{equation}\label{Eq:eq11}
F = \frac{\partial E_j}{\partial u} \text{ for } j = 1 \text{ and } 2,
\end{equation}
where \( F \) represents the reaction force corresponding to the prescribed displacement \( u \). The force-displacement curve for the single-unit cell (\(j=1\)), analyzed by numerically solving Eq.~\eqref{Eq:eq11} with the constraints given in Eq.~\eqref{Eq:eq3}, is depicted as the dashed black line in Fig.~\ref{fig:Fig2}(d). 
In contrast, for the two-unit cells (\(j = 2\)), we solve Eq.~\eqref{Eq:eq11} with the constraints in Eqs.~\eqref{Eq:eq7} and \eqref{Eq:eq8} using the same numerical methods and parameters. We obtain the force-displacement curve, depicted as the solid black line in Fig.~\ref{fig:Fig2}(d), which shows bistability. However, the prevalent von Mises truss model cannot capture the emergent bistability observed in Fig.~\ref{fig:Fig2}(c) and (d) (see Supplemental Material~\cite{SI} for details on the von Mises truss model).

\begin{figure}[t!]
    \centering
\includegraphics[width=1.0\linewidth]{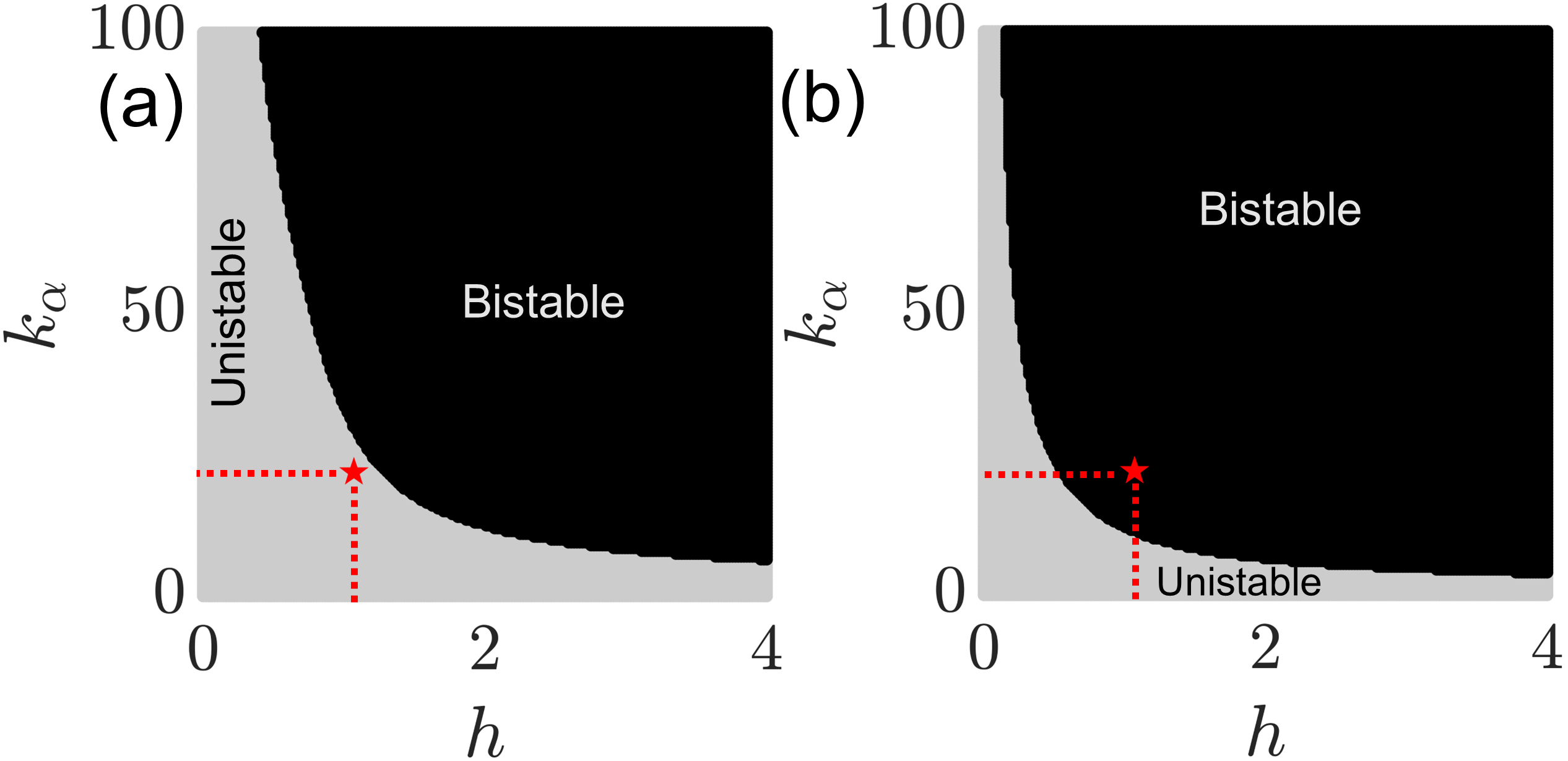}
    \caption{Parameter space of a single-unit cell and two-unit cell analytical discrete network: (a) demonstrates the parameter space of a single-unit cell by varying \(k_{\alpha}=\hat k_{\alpha}/\hat k_{\theta}\) and \(h=\hat{h}/\hat{b}\). The light gray zone indicates regions where a single-unit cell exhibits unistable behavior, whereas the black area denotes bistable behavior. (b) depicts the parameter space of two unit cell by varying \(k_{\alpha}=\hat k_{\alpha}/\hat k_{\theta}\) and \(h=\hat{h}/\hat{b}\). The red star \textcolor{red}{$\star$} denotes the parameters \((h,k_{\alpha}) = (1.07,22)\) used in the single-unit and two-unit cell study in Fig.~\ref{fig:Fig2}(c) and Fig.~\ref{fig:Fig2}(d).}
    \label{fig:Fig3}
\end{figure}
To design the mechanical metamaterial, Figs.~\ref{fig:Fig3}(a)-(b) illustrate the parameter spaces, taking into account the effect of \(k_{\alpha}\), and \(h\) for both single-unit cell and two-unit cell. These figures identify two regions: light gray for the unistable behavior and black for the bistable behavior. Two distinct stability regions are illustrated in Figs.~\ref{fig:Fig3}(a) and \ref{fig:Fig3}(b), characterized by the features of the force-displacement curve shown in Fig.~\ref{fig:Fig2}(d). The conditions are derived from Eq.~\eqref{Eq:eq11}, where \(\frac{\partial E_j}{\partial u} = F\) for \(j=1,2\). Unistability is identified when the minimum force \(F_{\min}\) remains positive (\(F_{\min} > 0\)). The corresponding results are depicted in the light gray region of Figs.~\ref{fig:Fig3}(a) and \ref{fig:Fig3}(b). Bistability is identified when the minimum force \(F_{\min}\) is negative (\(F_{\min} < 0\)). This behavior is represented in the black region of Figs.~\ref{fig:Fig3}(a) and \ref{fig:Fig3}(b). Figure~\ref{fig:Fig3}(a) depicts a 2D parameter space of \((k_{\alpha}, h)\) for a single-unit cell. As \(h\) and \(k_{\alpha}\) approaches zero, Fig.~\ref{fig:Fig3}(a) shows that the single unit moves towards unistable behavior. Conversely, as \(h\) and \(k_{\alpha}\) increase, the model shifts towards bistable behavior. Figure~\ref{fig:Fig3}(b) presents the 2D parameter space \((k_{\alpha}, h)\) for the two-unit cell discrete model. We observe a larger bistable area than the single-unit cell. Unlike Fig.~\ref{fig:Fig3}(a), Fig.~\ref{fig:Fig3}(b) demonstrates that the cooperative interactions between two-unit cells lead to bistable behavior at even lower \(h\) and \(k_{\alpha}\) values. Red star in Figs.~\ref{fig:Fig3}(a)-(b) denotes the parameters \((h,k_{\alpha}) = (1.07,22)\) used for single-unit and two-unit cell study in Figs.~\ref{fig:Fig2}(c)-(d).

\textit{Conclusion:} In conclusion, we reveal a new phenomenon: emergent bistability arising from cooperative interactions between two unistable unit cells, as demonstrated through experimental and finite element analysis. We also show that the conventional assumption of PBC overlooks the unit-cell interaction in mechanical metamaterials. Furthermore, our analytical discrete model captures the cooperative stiffening effect between two unistable unit cells and demonstrates emergent bistability, which the von Mises truss model fails to capture. This research opens new pathways for designing mechanical metamaterials with customizable stability suitable for adaptive and reconfigurable applications. \\

We acknowledge the support from the National Institutes of Health (NIH): Project No.\,R01EB032959. The theoretical part of this study is also partially funded by the National Science Foundation under Grant
No.\,2502227. Start-up funds from the Department of Mechanical Engineering at the Univ.\,of\,Utah also supported this work. The support and resources from the Center for High-Performance Computing at Univ.\,of\,Utah are gratefully acknowledged. The authors highly appreciate deep discussions with Profs, R. Parker and J. Hochhalter at Univ. of Utah.

\normalem
\bibliography{ref_NoTitle}      

\end{document}


\title{Supporting Information for \\
\emph{Rolling Waves with Non-Paraxial Phonon Spins}}

\author{Peng Zhang}
\affiliation{Department of Mechanical Engineering, University of Utah, Salt Lake City, UT 84112, USA}

\author{Christian Kern}
\affiliation{Department of Mathematics, University of Utah, Salt Lake City, UT 84112, USA}

\author{Sijie Sun}
\affiliation{Harvard John A. Paulson School of Engineering and Applied Science, Harvard University, Cambridge, MA 02138, USA}

\author{David A. Weitz}
\affiliation{Harvard John A. Paulson School of Engineering and Applied Science, Harvard University, Cambridge, MA 02138, USA}

\author{Pai Wang}%
\affiliation{Department of Mechanical Engineering, University of Utah, Salt Lake City, UT 84112, USA}
\thanks{pai.wang@utah.edu}

\maketitle


\title{Supporting Information for \\
\emph{Beyond von Mises Truss Models: Emergent Bistability in Mechanical Metamaterials\\}}

\maketitle
\noindent Md Nahid Hasan, Taylor E. Greenwood, Sharat Paul, Bolei Deng, Qihan Liu, Yong Lin Kong, and Pai Wang\\



\clearpage
\tableofcontents

\clearpage

\section{Geometric dimension of the single-unit cell}
\begin{figure}[htb!] 
    \centering    \includegraphics[width=0.70\linewidth]{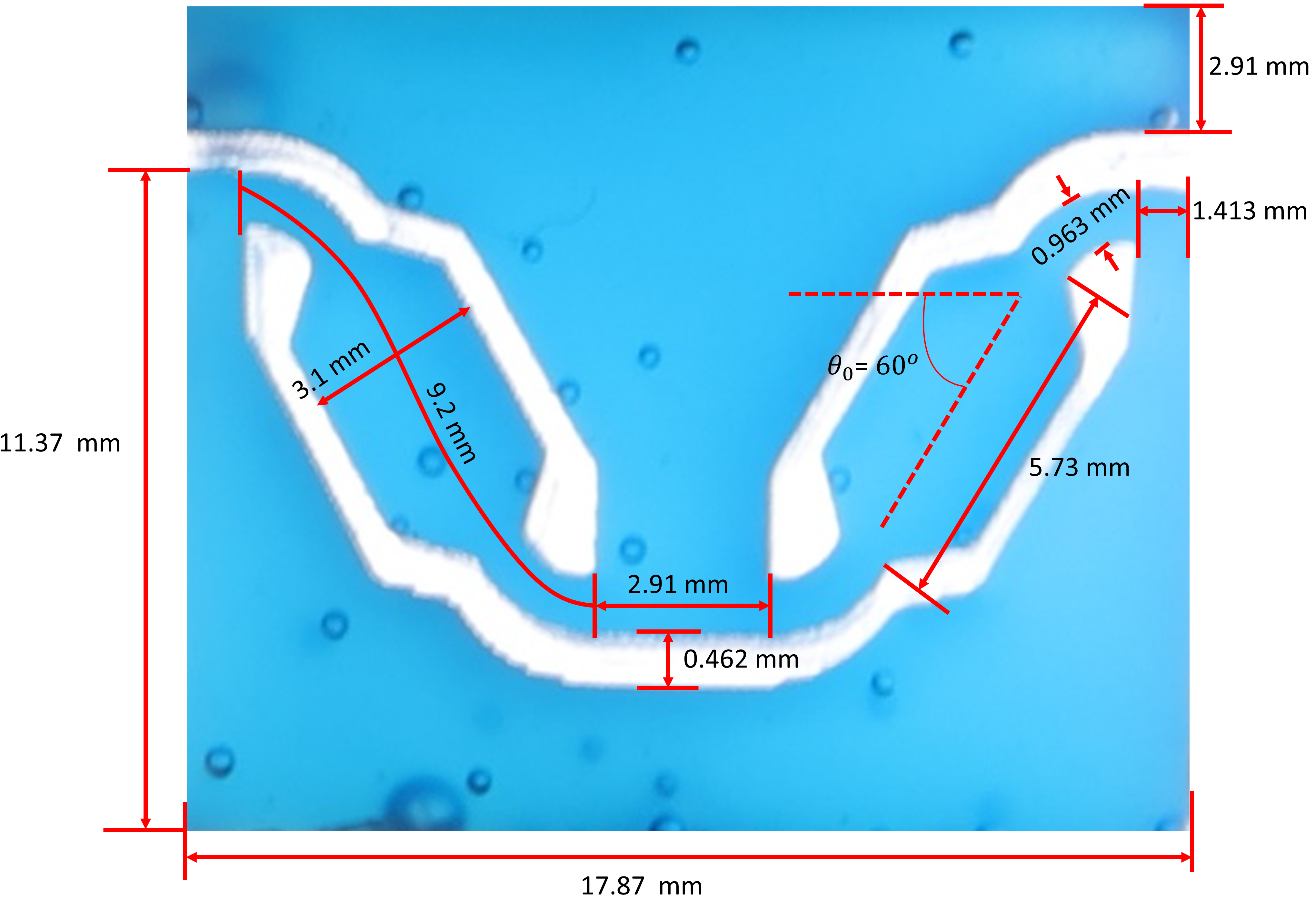}
    \caption{Geometric dimension of the mechanical metamaterial unit cell}
    \label{fig:Fig1} 
\end{figure}

Figure~\ref{fig:Fig1} shows the detailed geometric dimensions of the single-unit cell used in the main manuscript for both experimental and finite element analyses. 
\clearpage
\section{Hyperelastic material behavior}
\begin{figure}[htb!]
    \centering
    \includegraphics[width=0.55\linewidth]{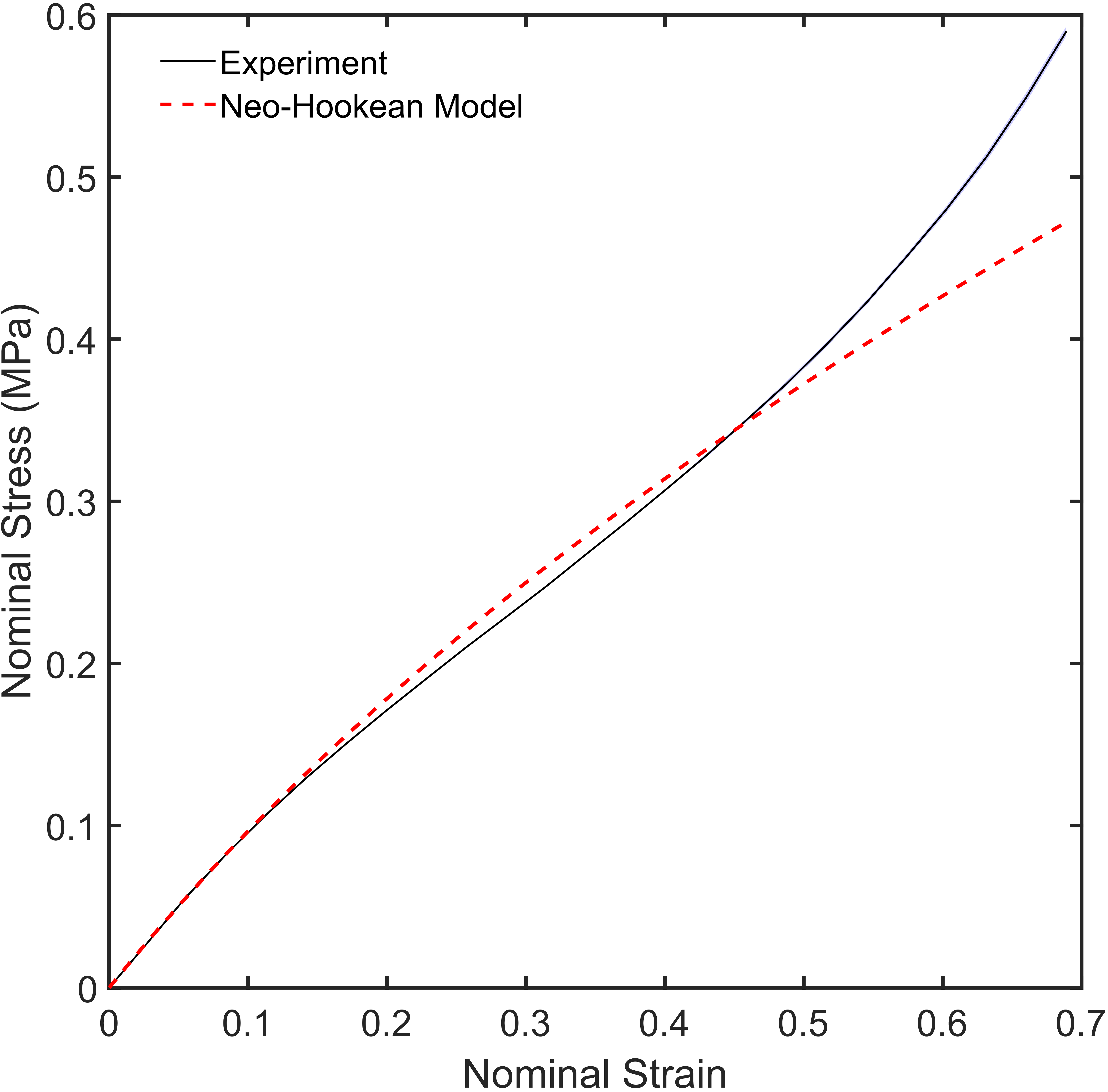}
    \caption{Nominal stress versus nominal strain under uniaxial tension for Dragon Skin 30. The solid black line shows the average of five experimental trials, and the red dashed line shows the Neo-Hookean model. The blue-shaded region indicates the standard deviation across the experiments.}
    \label{fig:Fig2}
\end{figure}

The hyperelastic material Dragon Skin 30~\cite{DragonSkin30} is modeled as nearly incompressible with Poisson's ratio of \(\nu=0.495\). The initial Young's modulus is estimated to be \(Y = 0.74\pm 0.07\) MPa, based on the initial slope of the uniaxial tensile test data depicted in Fig.~\ref{fig:Fig2}. The shear modulus \(G\) is derived from the relationship \(G = \frac{Y}{2(1+\nu)}\), and the bulk modulus \(K\) is calculated using \(K = \frac{Y}{3(1-2\nu)}\). Within the framework of the neo-Hookean model~\cite{shan2015multistable}, the primary material constant \(C_{10}\) is defined as half of the shear modulus \(C_{10} = \frac{G}{2}\), and the compressibility factor \(D_1\) is inversely proportional to twice the bulk modulus \(D_1 = \frac{2}{K}\)~\cite{AbaqusDoc2009}.

Using the neo-Hookean model, we utilize the aforementioned parameter values for the finite element analysis of single and two-unit cells.
\section{Analytical discrete model for single-unit cell}
\begin{figure}[htb!]
    \centering
    \includegraphics[width=0.45\linewidth]{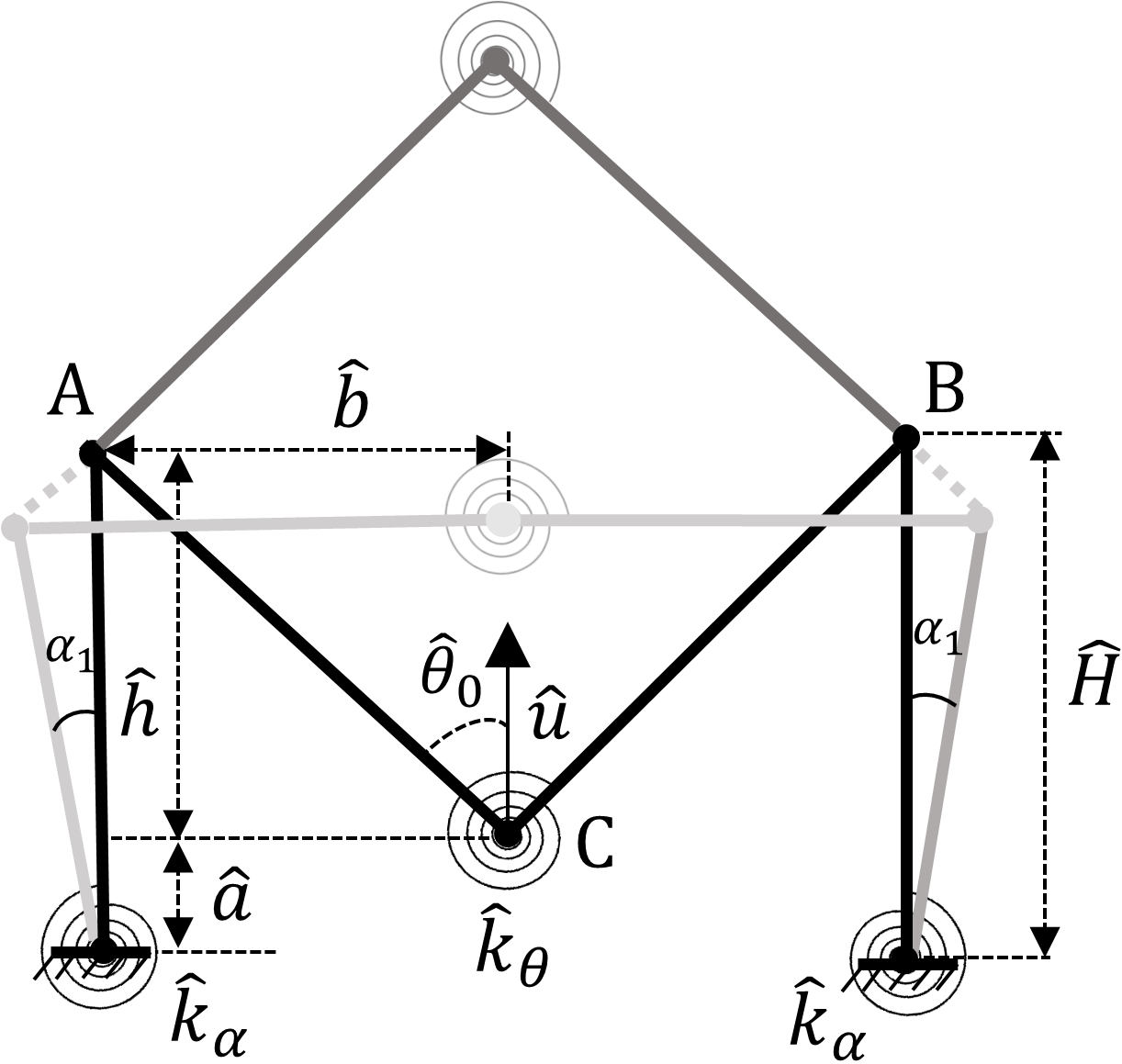}
    \caption{An analytical discrete model of a single-unit cell.}
    \label{fig:Fig3} %
\end{figure}

We present an analytical discrete model of a single unit cell as illustrated in Fig.~\ref{fig:Fig3}. The model comprises two middle rigid incline bars, denoted as \(\textrm{AC}\) and \(\textrm{BC}\), each with an initial length of \( \hat{L}_0 = \sqrt{\hat{b}^2 + \hat{h}^2} \). These bars incline at an angle \( \hat{\theta}_0 = \tan^{-1}\left(\frac{\hat{b}}{\hat{h}}\right) \) from the vertical. The bars are connected at their midpoint C, located at an apex height of \(\hat{h}\), with the endpoints A and B separated by a width of \(2\hat{b}\). A rotational spring at C, with stiffness \( \hat{k}_{\theta} \), provides bending resistance to bar AC and BC. Two vertical sidebars, each of height \( \hat{H} = \hat{h} + \hat{a} \), are connected to the edge points A and B. These sidebars are fixed at their bases and have a rotational stiffness of \( \hat{k}_{\alpha} \), serving to stabilize and support the inclined bars by limiting lateral and rotational movements. Applying a vertical displacement \(\hat{u}\) at midpoint C induces an inclination \(\alpha_1\) in the sidebars, resulting in both a lateral displacement \(\hat{H} \sin \alpha_1\) and a reduction in the vertical height of the incline bars to \(\hat{H} - \hat{H} \cos \alpha_1\). Fig.~\ref{fig:Fig3} depicts a one-degree-of-freedom system, driven by the displacement \(\hat{u}\). The deformed angle of the incline bars AC and BC is defined by:
\begin{equation} \label{Eq:S1}
\hat{\theta_0'} = \tan^{-1}\left(\frac{\hat{b} + \hat{H} \sin \alpha_1}{\hat{h} - \hat{H} - \hat{u} + \hat{H} \cos \alpha_1}\right).
\end{equation}
Additionally, the configuration must satisfy the geometric constraint ensuring the constant diagonal length of bars AC and BC, influenced by \(\alpha_1\), as described by:
\begin{equation} \label{Eq:S2}
\sqrt{\hat{b}^2 + \hat{h}^2}=\sqrt{(\hat{b} + \hat{H} \sin \alpha_1)^2 + (\hat{h} - \hat{H} - \hat{u} + \hat{H} \cos \alpha_1)^2}.
\end{equation}
This constraint confirms that the changes in sidebar inclination due to \(\hat{u}\) do not alter the overall length of bars AC and BC. Therefore, the total potential energy for a single-unit cell, \(\hat{E}_1\), is given by
\begin{equation} \label{Eq:S3}
\hat{E}_1 = \frac{1}{2}\hat{k}_{\theta} \left( 2\Delta \hat{\theta_0} \right)^2+\hat{k}_{\alpha} \alpha_{1}^2 , 
\end{equation}
where \(\Delta \hat{\theta_0} = (\hat \theta -\hat \theta_0')=\left( \tan^{-1}\left(\frac{\hat{b} + \hat{H} \sin \alpha_1}{\hat{h} - \hat{H} - \hat{u} + \hat{H} \cos \alpha_1}\right) - \tan^{-1}\left(\frac{\hat{b}}{\hat{h}}\right) \right)\) represents the change in angle of bars AC and BC under the prescribed displacement \(\hat{u}\).

\begin{align}
\hat{E}_1 = \frac{1}{2}\hat{k}_{\theta} \left( 2 \bigg( \tan^{-1} \bigg( \frac{\hat{b} + \hat{H} \sin \alpha_1}{\hat{h} - \hat{H} - \hat{u} + \hat{H} \cos \alpha_1} \bigg) - \tan^{-1} \bigg( \frac{\hat{b}}{\hat{h}} \bigg) \bigg) \right)^2 + \hat{k}_{\alpha} \alpha_{1}^2.
\label{Eq:S4}
\end{align}

In order to simplify Eq.~\eqref{Eq:S4}, we perform normalization with respect to \(\hat k_{\theta}\) and \(\hat b\), setting both \(\hat k_{\theta} = 1\) and \(\hat b = 1\). The resulting dimensionless variables are expressed as follows:
\begin{align}
    u &= \frac{\hat{u}}{\hat{b}}, \quad h = \frac{\hat{h}}{\hat{b}}, \quad H = \frac{\hat{H}}{\hat{b}}, \quad a = \frac{\hat{a}}{\hat{b}}, \quad {E_1} = \frac{\hat{E}_1}{\hat{k}_{\theta}}, \quad k_{\alpha} = \frac{\hat{k}_{\alpha}}{\hat{k}_{\theta}}.
    \label{Eq:S5}
\end{align}
Accordingly, Eq.~\eqref{Eq:S4} becomes Eq.~\eqref{Eq:S6}:
\begin{equation} \label{Eq:S6}
E_1 = \frac{1}{2}\left( 2\Delta \theta_0 \right)^2+k_{\alpha} \alpha_{1}^2,
\end{equation}
With the geometric constraint given by
\begin{equation} \label{Eq:S7}
\sqrt{1 + h^2}=\sqrt{(1 + H \sin \alpha_1)^2 + (h - H - u + H \cos \alpha_1)^2},
\end{equation}
and change the angle of the two inclined bars under displacement \(u\)
\begin{equation} \label{Eq:S8}
\Delta \theta_0 = \left( \tan^{-1}\left(\frac{1+H \sin \alpha_1}{h - H - u + H \cos \alpha_1}\right) - \tan^{-1}\left(\frac{1}{h}\right) \right).
\end{equation}
We numerically solve Eq.~\eqref{Eq:S6}  using the arc-length method with the geometric constraint of the two inclined bars to obtain the energy landscape for the main manuscript's single-unit cell described in Fig. 2(c).

\section{Analytical discrete model for two-unit cell}

\begin{figure}[htb!]
    \centering
    \includegraphics[width=0.75\linewidth]{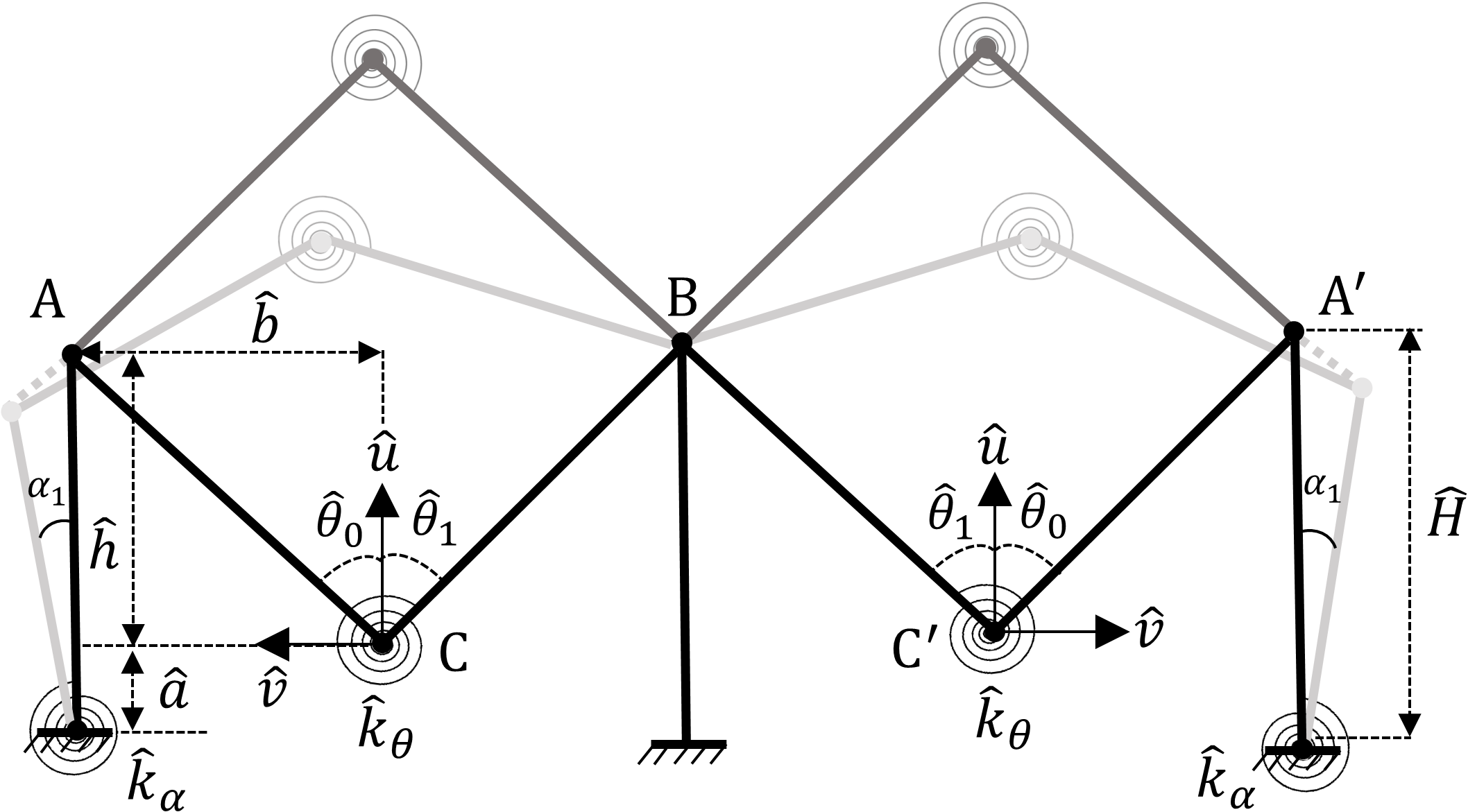}
    \caption{An analytical discrete model of a two-unit cell.}
    \label{fig:Fig4}
\end{figure}

Building on the analytical discrete model depicted in Fig.~\ref{fig:Fig3}, we extend the configuration to a two-unit cell network, shown in Fig.~\ref{fig:Fig4}. This configuration introduces interactions between the units but does not add independent degrees of freedom. Instead, the horizontal displacement \( \hat{v} \) and the rotation angle \( \alpha_1 \) of the sidebars are dependent variables, determined by the vertical displacement \( \hat{u} \) and the geometric constraints of the system. Fig.~\ref{fig:Fig4} also shows that the inclination angles of bars AC (and, by symmetry, also \(\textrm{A}'\textrm{C}'\)) and \(\textrm{B}\textrm{C}\) (by symmetry, also \(\textrm{B}\textrm{C}'\)) are denoted as \({\theta}_0 = {\theta}_1 = \tan^{-1}\left(\frac{\hat{b}}{\hat{h}}\right)\). The model, as illustrated in Fig.~\ref{fig:Fig4}, effectively remains a one-degree-of-freedom system, driven by the displacement \( \hat{u} \).
The governing geometric constraints for the inclined bars are defined as follows:
\begin{equation}
\label{Eq:S9}
\sqrt{\hat{b}^2 + \hat{h}^2}=\sqrt{(\hat{b} - \hat{H} \sin \alpha_1 - \hat{v})^2 + (\hat{h} - \hat{H} - \hat{u} + \hat{H} \cos \alpha_1)^2},
\end{equation}

for the side incline bars such as \( \textrm{AC} \) and \( \textrm{A}'\textrm{C}' \), and
\begin{equation}
\label{Eq:S10}
\sqrt{\hat{b}^2 + \hat{h}^2}=\sqrt{(\hat{b} + \hat{v})^2 + (\hat{h} - \hat{u})^2},
\end{equation}
for the middle incline bars such as \( \textrm{BC} \) and \( \textrm{BC}' \).
The total potential energy for the two-unit cell is denoted by \( \hat{E}_2 \) and is given by:
\begin{equation}
\label{Eq:S11}
\hat{E}_2 =\hat{k}_{\theta} \left( \Delta\hat{\theta_{0}}+ \Delta \hat{\theta_{1}} \right)^2+ \hat{k}_{\alpha} \alpha_1^2,
\end{equation}
where \( \Delta \hat{\theta_0}=(\hat{\theta_0}'-\hat{\theta_0} \)) and \( \hat{\theta_1}= (\hat{\theta_1}'- \hat{\theta_1} \)) is defined as:
\begin{equation}
\label{Eq:S12}
\Delta \hat{\theta_0} =\left( \tan^{-1}\left(\frac{\hat{b} + \hat{H} \sin \alpha_1-v}{\hat{h} - \hat{H} - \hat{u} + \hat{H} \cos \alpha_1}\right) - \tan^{-1}\left(\frac{\hat{b}}{\hat{h}}\right) \right) \textrm{for bar AC} 
 ~\textrm{and}~\textrm{A}'\textrm{C}',
\end{equation}
\begin{equation}
\label{Eq:S13}
\Delta \hat{\theta_1} = \left( \tan^{-1}\left(\frac{\hat{b} + \hat{v}}{\hat{h} - \hat{u}}\right) - \tan^{-1}\left(\frac{\hat{b}}{\hat{h}}\right) \right)  \textrm{for bar BC} 
 ~\textrm{and}~\textrm{B}\textrm{C}'.
\end{equation}
Note that, due to prescribed vertical displacement \(u\), inclination \(\alpha_1\) of the sidebars, and horizontal translation \(v\), the deformed angles of bar AC (by symmetry, also \(\textrm{A}'\textrm{C}'\)) and \(\textrm{B}\textrm{C}\) (by symmetry, also \(\textrm{B}\textrm{C}'\)) are denoted as \(\hat{\theta}_0' = \tan^{-1}\left(\frac{\hat{b} + H \sin \alpha_1 - \hat{v}}{\hat{h} - H - \hat{u} + H \cos \alpha_1}\right)\) and \(\hat{\theta}_1' = \tan^{-1}\left(\frac{\hat{b} + \hat{v}}{\hat{h} - \hat{u}}\right)\), respectively. Normalization of Eq.~\eqref{Eq:S11} with respect to \(\hat{k}_{\theta}\) and \(\hat{b}\) involves setting both \(\hat{k}_{\theta} = 1\) and \(\hat{b} = 1\), leading to the following dimensionless variables:
\begin{align}
\label{Eq:S14}
u &= \frac{\hat{u}}{\hat{b}}, \quad h = \frac{\hat{h}}{\hat{b}}, \quad H = \frac{\hat{H}}{\hat{b}}, \quad a = \frac{\hat{a}}{\hat{b}}, \quad E_2 = \frac{\hat{E}_2}{\hat{k}_{\theta}}, \quad k_{\alpha} = \frac{\hat{k}_{\alpha}}{\hat{k}_{\theta}}.
\end{align}
Thus, Eq.~\eqref{Eq:S11} becomes:
\begin{equation}
\label{Eq:S15}
E_2 =\left( \Delta \theta_0+\Delta \theta_1 \right)^2+ k_{\alpha} \alpha_1^2,
\end{equation} 
With geometric constraint
\begin{equation}
\label{Eq:S16}
\sqrt{1^2 + h^2}=\sqrt{(1 - H \sin \alpha_1 - v)^2 + (h - H - u + H \cos \alpha_1)^2},
\end{equation}
\begin{equation}
\label{Eq:S17}
\sqrt{1^2 + h^2}=\sqrt{(1 + v)^2 + (h - u)^2},
\end{equation}

where \(\Delta \theta_0\)  and \(\Delta \theta_1\) are denoted as:
\begin{equation} \label{Eq:S18}
\Delta \theta_0 = \left( \tan^{-1}\left(\frac{1+H \sin \alpha_1-v}{h - H - u + H \cos \alpha_1}\right) - \tan^{-1}\left(\frac{1}{h}\right) \right) \textrm{for bar AC} 
 ~\textrm{and}~\textrm{A}'\textrm{C}',
\end{equation}
\begin{equation}
\label{Eq:S19}
\Delta \theta_1 = \left( \tan^{-1}\left(\frac{1+v}{h - u}\right) - \tan^{-1}\left(\frac{1}{h}\right) \right) \textrm{for bar BC} 
 ~\textrm{and}~\textrm{B}\textrm{C}'.
\end{equation}
The energy landscape for this two-unit cell system is obtained numerically using the arc-length method, incorporating the geometric constraints of the inclined bars as described in Fig. 2(c) of the main manuscript.
\section{von Mises Truss model}
We demonstrate that the traditional von Mises Truss model cannot capture the cooperative interactions between two unit cells~\cite{yang2023shape}. With certain parameters, a single-unit cell is unistable. However, even when an identical analytical spring model is added, it still exhibits unistability.
\subsection{Single-unit cell}

\begin{figure}[t!]
    \centering
\includegraphics[width=0.35\linewidth]{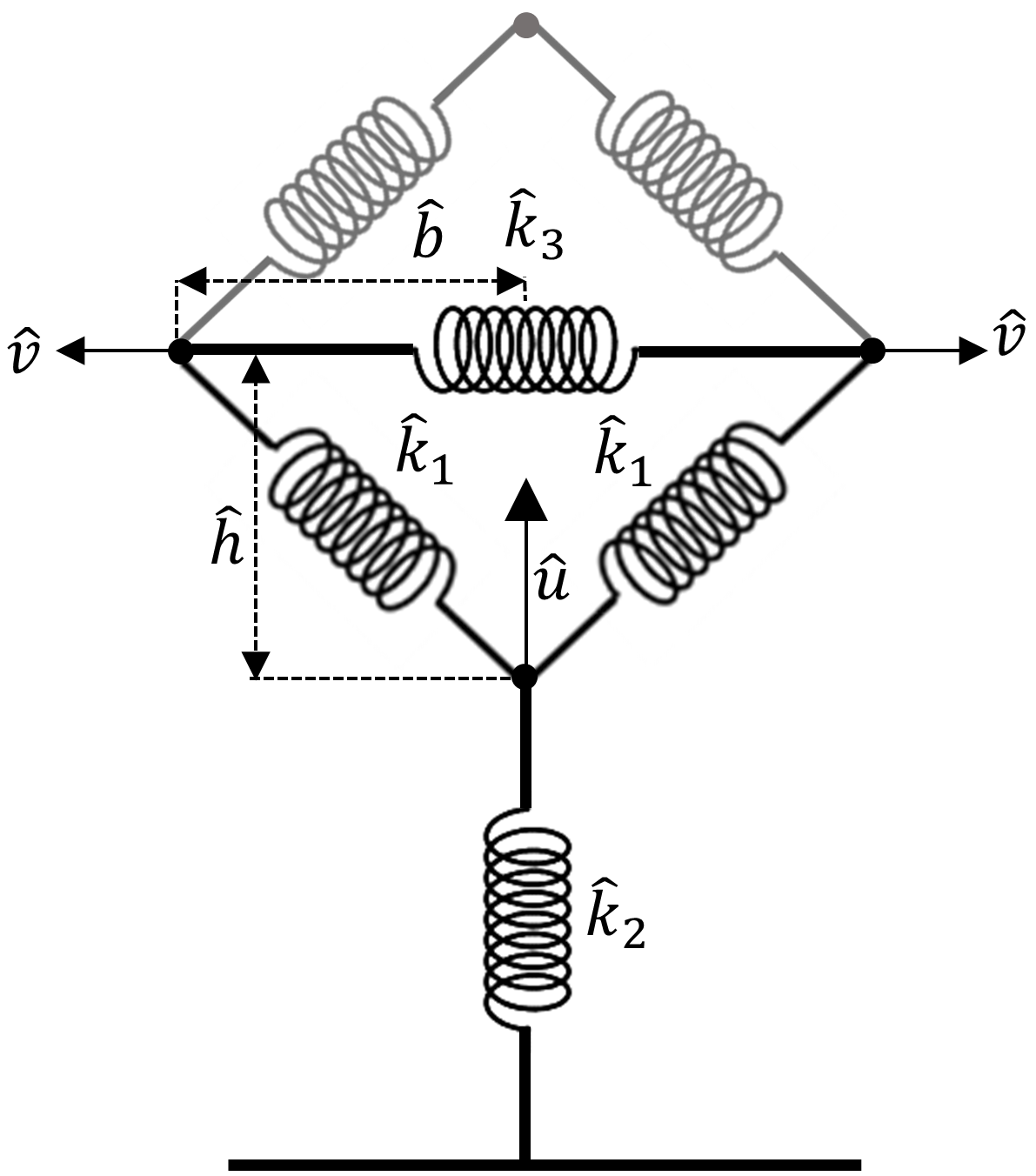}
    \caption{von Mises Truss model for a single-unit cell.}
    \label{fig:Fig5}
\end{figure}

When the vertical controlled displacement \(\hat{u}\) or force \(\hat{F}\) is applied to the single-unit cell of Fig.~\ref{fig:Fig5}, the conservative system's total potential energy, \(E_1\), consists of the potential elastic energy from two inclined springs, horizontal and vertical springs, and the work done by the force~\cite{yang2023shape}. The total potential energy of the single-unit cell is given by Eq.~\eqref{Eq:S20}:

\begin{equation} \label{Eq:S20}
\hat{E}_1 = \frac{\hat{k}_{2} \hat{u}^2}{2} + 2 \hat{k}_{3} \hat{v}^2 + \hat{k}_{1} \left(\sqrt{\hat{b}^2 + \hat{h}^2} - \sqrt{(\hat{b} - \hat{v})^2 + (\hat{h} - \hat{u})^2}\right)^2 - \hat{F} \hat{u}.
\end{equation}

The equilibrium path of Eq.~\eqref{Eq:S20} is determined through the stationary condition of the total potential energy, i.e., \(\delta \hat{E}_1 = 0\). The force-displacement curve of the analytical spring network is obtained by numerically solving Eq.~\eqref{Eq:S21}.

\begin{equation}\label{Eq:S21}
\frac{\partial \hat{E}_1}{\partial \hat{u}} = \hat{F}, \quad \frac{\partial \hat{E}_1}{\partial \hat{v}} = 0.
\end{equation}

\begin{align}\label{Eq:S22}
\hat{F} = \hat{k}_{2}\,\hat{u} + \frac{2\hat{k}_{1}\,\left(\sqrt{\hat{b}^2 + \hat{h}^2} - \sqrt{{\left(\hat{b} - \hat{v}\right)}^2 + {\left(\hat{h} - \hat{u}\right)}^2}\right)\,\left(\hat{h} - \hat{u}\right)}{\sqrt{{\left(\hat{b} - \hat{v}\right)}^2 + {\left(\hat{h} - \hat{u}\right)}^2}}
\end{align}

\begin{align}\label{Eq:S23}
4\,\hat{k}_{3}\,\hat{v} + \frac{2\hat{k}_{1}\,\left(\sqrt{\hat{b}^2 + \hat{h}^2} - \sqrt{{\left(\hat{b} - \hat{v}\right)}^2 + {\left(\hat{h} - \hat{u}\right)}^2}\right)\,\left(\hat{b} - \hat{v}\right)}{\sqrt{{\left(\hat{b} - \hat{v}\right)}^2 + {\left(\hat{h} - \hat{u}\right)}^2}} = 0
\end{align}

We normalize the model's dimensions relative to \( \hat{b} \), and the force, stiffness, and energy terms by \( \hat{k}_{1} \), by setting \( \hat{b} = 1 \) and \( \hat{k}_{1} = 1 \), resulting in dimensionless parameters. This normalization yields \( u = \frac{\hat{u}}{\hat{b}} \), \( h = \frac{\hat{h}}{\hat{b}} \), \( E_1 = \frac{\hat{E}_1}{\hat{k}_{1} \hat{b}^2} \), \( F = \frac{\hat{F}}{\hat{k}_{1} \hat{b}} \), \( k_{3} = \frac{\hat{k}_{3}}{\hat{k}_{1} \hat{b}^2} \), and \( k_{2} = \frac{\hat{k}_{2}}{\hat{k}_{1} \hat{b}^2} \). Therefore the  Eq.~\eqref{Eq:S22} and  Eq.~\eqref{Eq:S23} becomes,

\begin{align}\label{Eq:S24}
F = k_{2}\,u + \frac{2\left(\sqrt{1 + h^2} - \sqrt{(1 - v)^2 + (h - u)^2}\right)\,(h - u)}{\sqrt{(1 - v)^2 + (h - u)^2}},
\end{align}

\begin{align}\label{Eq:S25}
4\,k_{3}\,v + \frac{2\left(\sqrt{1 + h^2} - \sqrt{(1 - v)^2 + (h - u)^2}\right)\,(1 - v)}{\sqrt{(1 - v)^2 + (h - u)^2}} = 0.
\end{align}

\subsection{Two-unit cell}
\begin{figure}[t!]
    \centering
\includegraphics[width=0.5\linewidth]{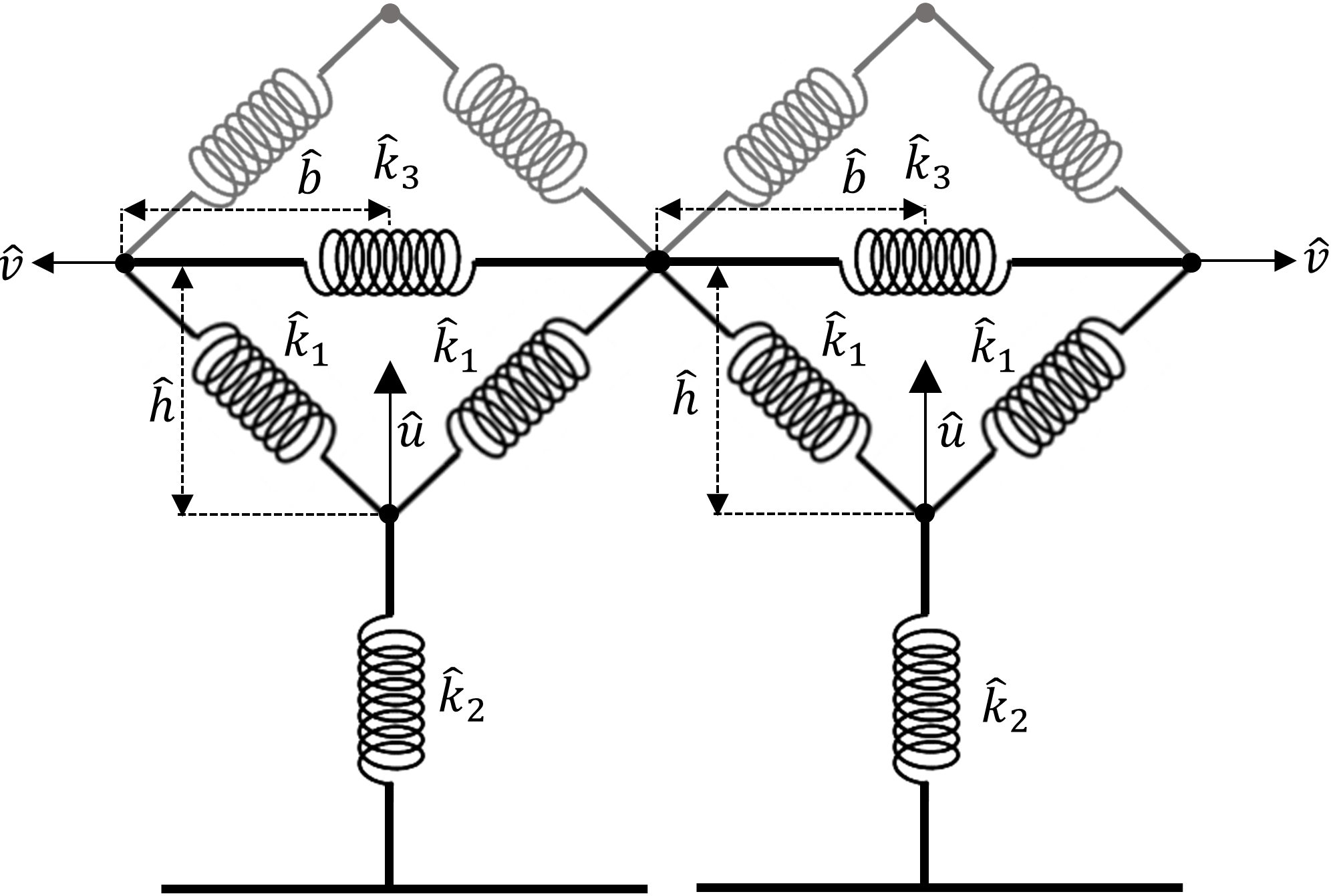}
    \caption{von Mises Truss model for a two-unit cell.}
    \label{fig:Fig6}
\end{figure}

When the vertical controlled displacement \(\hat{u}\) or force \(\hat{F}\) is applied to the two-unit cell of Fig.~\ref{fig:Fig6}, the conservative system's total potential energy, \(E_2\), consists of the potential elastic energy from two inclined springs, horizontal and vertical springs, and the work done by the force~\cite{yang2023shape}. The total potential energy of the two-unit cell is given by Eq.~\eqref{Eq:S26}:

\begin{align}\label{Eq:S26}
\hat{E}_2 = &\hat{k}_{2}\,\hat{u}^2 - \hat{F}\,\hat{u} + \hat{k}_{3}\,\hat{v}^2 \nonumber \\
&+ 2\hat{k}_{1}\left(\sqrt{\hat{b}^2 + \hat{h}^2} - \sqrt{(\hat{b} - \hat{v})^2 + (\hat{h} - \hat{u})^2}\right)^2 \nonumber \\
\end{align}

The equilibrium path of Eq.~\eqref{Eq:S26} is determined through the stationary condition of the total potential energy, i.e., \(\delta \hat{E}_2 = 0\). The force-displacement curve of the analytical spring network is evaluated by numerically solving Eq.~\eqref{Eq:S27}:
\begin{equation}\label{Eq:S27}
\frac{\partial \hat{E}_2}{\partial \hat{u}} = \hat{F}, \quad \frac{\partial \hat{E}_2}{\partial \hat{v}} = 0.
\end{equation}

\begin{align}\label{Eq:S28}
\hat{F} = &2\hat{k}_{2}\hat{u} + \frac{4\hat{k}_{1}\left(\sqrt{\hat{b}^2 + \hat{h}^2} - \sqrt{(\hat{b} - \hat{v})^2 + (\hat{h} - \hat{u})^2}\right)(\hat{h} - \hat{u})}{\sqrt{(\hat{b} - \hat{v})^2 + (\hat{h} - \hat{u})^2}}= 0
\end{align}

\begin{align}\label{Eq:S29}
2\hat{k}_{3}\hat{v} + \frac{4\hat{k}_{1}\left(\sqrt{\hat{b}^2 + \hat{h}^2} - \sqrt{(\hat{b} - \hat{v})^2 + (\hat{h} - \hat{u})^2}\right)(\hat{b} - \hat{v})}{\sqrt{(\hat{b} - \hat{v})^2 + (\hat{h} - \hat{u})^2}} = 0
\end{align}

\begin{figure}[t!]
    \centering
    \includegraphics[width=0.5\linewidth]{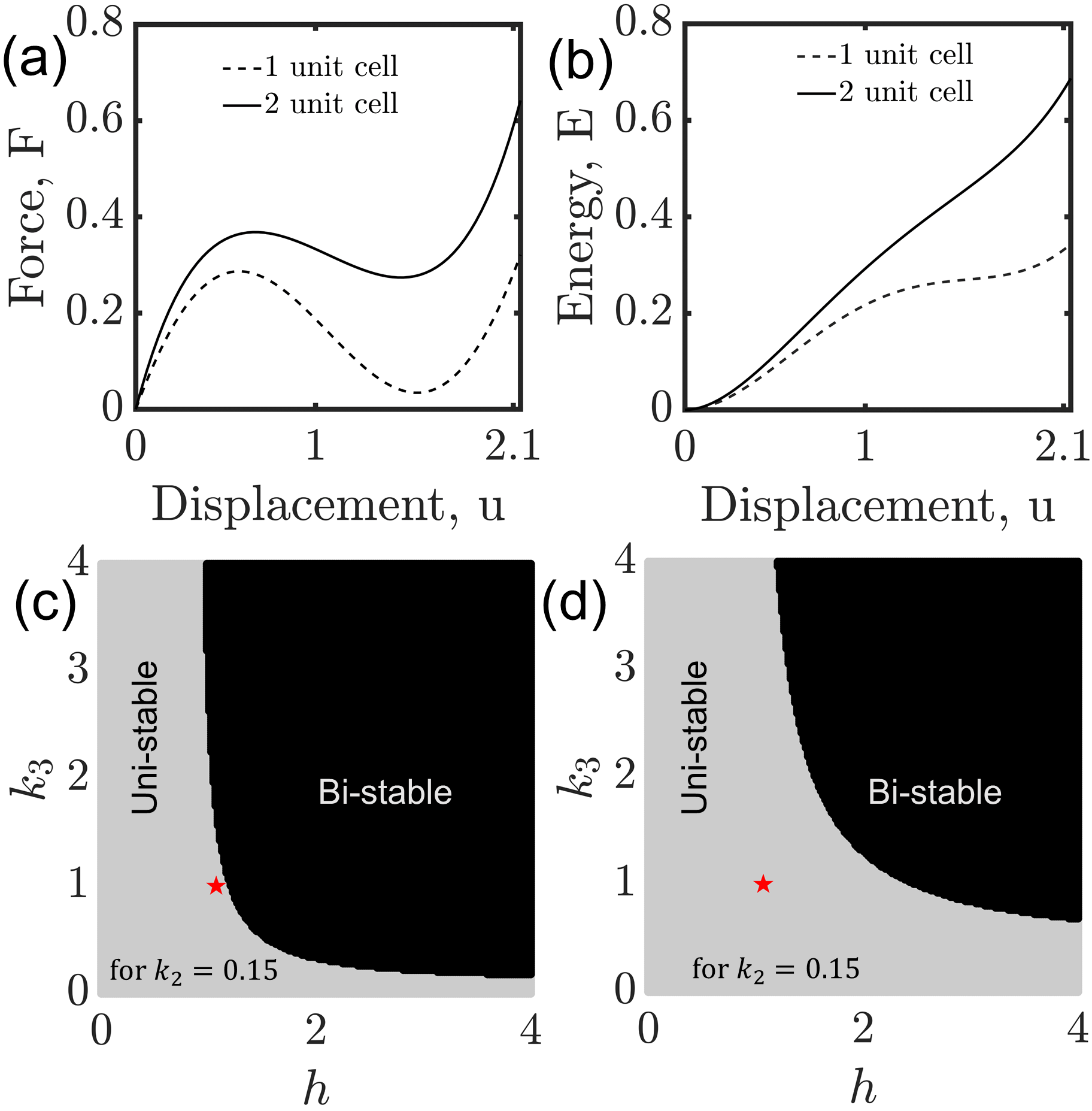}
    \caption{von Mises Truss model and its limitations: (a) Force-displacement curves for single and two-unit cells showing unistable behavior. The dashed line denotes the single-unit cell, and the solid line denotes the two-unit cell, with parameters \((h, k_2, k_3) = (1.07, 0.15, 1)\). (b) Energy landscapes for single-unit and two-unit cells, with the same parameter settings. There are no distinct energy barriers for either cell type. (c) Parameter space for single-unit and two-unit cells, varying \(k_3\) and \(h\) for \(k_2 = 0.15\). The light gray region indicates unistable behavior, while the black area indicates bistable behavior. The red star marks the parameters \((h, k_3) = (1.07, 1)\) used in the study.}
    \label{fig:Fig7}
\end{figure}

We normalize the model's dimensions relative to \( \hat{b} \) and the force, stiffness and energies by \( \hat{k}_{1} \), by setting the values of \( \hat{b} \) and \( \hat{k}_{1} \) to 1, resulting in dimensionless parameters. This normalization yields \( u = \frac{\hat{u}}{\hat{b}} \), \( h = \frac{\hat{h}}{\hat{b}} \), \( E_2 = \frac{\hat{E}_1}{\hat{k}_{1} \hat{b}^2} \), \( F = \frac{\hat{F}}{\hat{k}_{1} \hat{b}} \), \( k_{3} = \frac{\hat{k}_{3}}{\hat{k}_{1} \hat{b}^2} \), and \( k_{2} = \frac{\hat{k}_{2}}{\hat{k}_{1} \hat{b}^2} \). Therefore the  Eq.~\eqref{Eq:S28} and  Eq.~\eqref{Eq:S29} becomes,

\begin{align}\label{Eq:S30}
F=2k_{2}\,u+\frac{4k_{1}\,\left(\sqrt{1+h^2}-\sqrt{{\left(1-v\right)}^2+{\left(h-u\right)}^2}\right)\,\left(h-u\right)}{\sqrt{{\left(1-v\right)}^2+{\left(h-u\right)}^2}}=0
\end{align}

\begin{align}\label{Eq:S31}
2k_{3}\,v+\frac{4k_{1}\,\left(\sqrt{1+h^2}-\sqrt{{\left(1-v\right)}^2+{\left(h-u\right)}^2}\right)\,\left(1-v\right)}{\sqrt{{\left(1-v\right)}^2+{\left(h-u\right)}^2}}=0
\end{align}
Fig.~\ref{fig:Fig7}(a) shows the force-displacement curve for single-unit and two-unit cells obtained by numerically solving Eqs.~\eqref{Eq:S24} and \eqref{Eq:S25} for the single-unit cell and Eqs.~\eqref{Eq:S30} and \eqref{Eq:S31} for the two-unit cell. For the single-unit cell, we first solve Eq.~\eqref{Eq:S25} numerically for prescribed values of \(u\) to find corresponding \(v\) values. These \(v\) values, along with the prescribed \(u\) values, are then used to solve Eq.~\eqref{Eq:S24} and obtain the force-displacement curve. A similar numerical approach is used for the two-unit cell with Eqs.~\eqref{Eq:S30} and \eqref{Eq:S31}. By numerically integrating the force-displacement curve in Fig.~\ref{fig:Fig7}(a), we obtain the energy landscape for the single-unit and two-unit cells, as shown in Fig.~\ref{fig:Fig7}(b). Next, by following the same stability criteria as for single-unit and two-unit cells in the main manuscript, we demonstrate the parameter space of the von Mises Truss model by evaluating the force-displacement curve in Fig.~\ref{fig:Fig7}(a). Figs.~\ref{fig:Fig7}(c) and \ref{fig:Fig7}(d) show the parameter space for single-unit and two-unit cells, respectively. The red star in Figs.~\ref{fig:Fig7}(c) and \ref{fig:Fig7}(d) denotes the parameters \((h, k_{3}) = (1.07, 1)\) used for the single-unit and two-unit cell study in Figs.~\ref{fig:Fig7}(a) and \ref{fig:Fig7}(b) for \(k_2=0.15\). Unlike in the main manuscript, the parameter space in Figs.~\ref{fig:Fig7}(c) and \ref{fig:Fig7}(d) shows that the red star is in the unistable behavior zone for both the single-unit and two-unit cells.

\newpage
\bibliography{ref_NoTitle}
